\documentclass[aps,prd,nofootinbib,twocolumn,groupaddress,preprintnumbers]{revtex4-1}

\usepackage{amssymb}
\usepackage{amsmath}
\usepackage{epsfig}
\usepackage{hyperref}
\usepackage{breakurl}
\usepackage{bm}
\usepackage{color}
\usepackage{tikz}

\makeatletter
\def\simgt{\mathrel{\lower2.5pt\vbox{\lineskip=0pt\baselineskip=0pt
           \hbox{$>$}\hbox{$\sim$}}}}
\def\simlt{\mathrel{\lower2.5pt\vbox{\lineskip=0pt\baselineskip=0pt
           \hbox{$<$}\hbox{$\sim$}}}}

\makeatother

\def\spa#1.#2{\left\langle#1\,#2\right\rangle}
\def\spb#1.#2{\left[#1\,#2\right]}
\def\sand#1.#2.#3{%
\left\langle#1{\vphantom1}\right|{#2}\left|#3\right]}%
\def\sandmp#1.#2.#3{%
\left\langle#1{\vphantom1}\right|{#2}\left|#3\right]}%
\def\sandpm#1.#2.#3{%
\left[#1{\vphantom1}\right|{#2}\left|#3\right\rangle}%
\def\sandmm#1.#2.#3{%
\left\langle#1{\vphantom1}\right|{#2}\left|#3\right\rangle}%
\def\sandpp#1.#2.#3{%
\left[#1{\vphantom1}\right|{#2}\left|#3\right]}%

\def\Section#1{\vskip .05 cm \noindent {\it #1}}
\newcommand{\be}{\begin{equation}}
\newcommand{\ee}{\end{equation}}
\newcommand{\eq}[2]{\be\begin{aligned}#1 \label{#2}\end{aligned}\ee}

\newcommand{\Fig}[1]{Fig.~\ref{#1}}
\newcommand{\Eq}[1]{Eq.~\eqref{#1}}

\newcommand{\mbf}[1]{\boldsymbol{#1}}

\def\topbotatom#1{\hbox{\hbox to 0pt{$#1\bot$\hss}$#1\top$}}

\begin{document}

\title{Tidal Effects in the Post-Minkowskian Expansion}
\author{Clifford Cheung}
\author{Mikhail P. Solon}
\affiliation{Walter Burke Institute for Theoretical Physics,
    California Institute of Technology, Pasadena, CA 91125}
    
\begin{abstract}
Tools from scattering amplitudes and effective field theory have recently been repurposed to derive state-of-the-art results for the black hole binary inspiral in the post-Minkowskian expansion.  In the present work we extend this approach to include the tidal effects of mass and current quadrupoles on the conservative dynamics of non-spinning neutron star mergers. We compute the leading and, for the first time, next-to-leading order post-Minkowskian finite size corrections to the conservative Hamiltonian, together with their associated scattering amplitudes and scattering angles. Our expressions are gauge invariant and, in the extreme mass ratio limit, consistent with the dynamics of a tidally deformed test body in a Schwarzschild background.  Furthermore, they agree completely with existing results at leading post-Minkowskian and second post-Newtonian orders.

\end{abstract}

\preprint{CALT-TH 2020-025 }

\maketitle

\Section{Introduction.} The monumental discovery of gravitational waves by LIGO/Virgo~\cite{LIGO} has sparked a flurry of activity in applying ideas from the study of scattering amplitudes and effective field theory (EFT) to the binary inspiral problem. Building on seminal work on the quantum field theoretic description of gravitons~\cite{oldQFT,NRGR}, this nascent program has fused cutting edge tools from the double copy~\cite{DoubleCopy}, generalized unitarity~\cite{GeneralizedUnitarity}, and EFT~\cite{Neill:2013wsa, 2PM} to obtain the now state-of-the-art ${\cal O}(G^3)$ conservative Hamiltonian for spinless binary black holes~\cite{3PM, 3PMlong, 3PMFeynman}.  

New results have also been derived for binary systems with spin~\cite{newSPIN,Vines:2018gqi}, with supersymmetry~\cite{N8}, and for scattering of massless particles with or without supersymmetry~\cite{GravitonScattering}. Other advances in this area have utilized classic tools from quantum mechanics and quantum field theory~\cite{LSE, HBET}, newly uncovered amplitudes structures~\cite{newAMP}, and analytic continuation between the scattering and bound state problems~\cite{B2B}.  

There is, of course, an illustrious record of tackling this subject with conventional methods such as effective one-body formalism~\cite{EOB}, numerical relativity~\cite{NR}, the
self-force formalism~\cite{self_force}, and perturbative analysis
using post-Newtonian (PN)~\cite{PN}, post-Minkowskian (PM)~\cite{PM}, and
non-relativistic general relativity~\cite{NRGR,NRGR2}.   

Importantly, these more traditional approaches have all been adapted to a principal scientific aim of the gravitational wave program: disentangling the underlying nuclear properties of neutron stars (see Refs.~\cite{Buonanno:2014aza,Dietrich:2020eud} for reviews). Recent detections of gravitational waves generated by the inspiral and merger of neutron stars have already put direct constraints on the equation of state of matter at nuclear densities~\cite{TheLIGOScientific:2017qsa,Abbott:2020uma}, and much effort has been dedicated to the prospects and challenges for maximizing the science yield from current and future measurements~\cite{Flanagan:2007ix, MeasuringTidal}.
Tidal effects have been modeled using a variety of numerical~\cite{NRTidal} and analytic methods~\cite{Flanagan:2007ix, AnalyticTidal,Steinhoff:2016rfi,Henry:2019xhg}, including the self-force~\cite{SFTidal} and effective one-body~\cite{EOBTidal} formalisms, and very recently using PM perturbation theory~\cite{Bini:2020flp,PMGR}. These, together with the prospect of accurate measurements of tidal parameters at future third-generation detectors like the Einstein Telescope~\cite{ET}, all offer strong motivations to extend the tools of scattering amplitudes and EFT to incorporate  the corresponding finite size effects.

In this paper we compute the leading and next-to-leading PM conservative Hamiltonian induced by the mass and current quadrupole moments of spinless compact bodies.  
To begin, we compute Feynman diagrams describing the scattering of non-minimally coupled, gravitationally interacting massive scalars at one- and two-loop orders in a general field basis and gauge fixing.  These objects encode the leading ${\cal O}(G^2)$ and next-to-leading ${\cal O}(G^3)$ PM tidal corrections. We then integrate these Feynman diagrams via the methods of~\cite{2PM,3PMlong} to obtain a gauge invariant scattering amplitude.  Equating this to the amplitude computed in a non-relativistic EFT, we then derive the corresponding conservative Hamiltonian for tidal effects. Throughout, we work at linear order in the tidal coefficients while resumming to all orders in the velocity expansion.

As a check we compute the associated scattering angle and find exact agreement with the leading order PM results of \cite{Bini:2020flp,PMGR}.  At the relevant overlapping orders, our expressions are also consistent with existing results for the 2PN Hamiltonian and 1PN binding energy~\cite{Henry:2019xhg,EOBTidal}.  Last but not least, the test-particle limit of our Hamiltonian is gauge equivalent to that of a tidally deformed test mass in a Schwarzschild background.  

\medskip

\Section{Setup.} Our setup is described by a pair of massive scalars interacting gravitationally via minimal coupling\footnote{We work in mostly plus metric signature throughout.}
\eq{
S& =  \int d^4 x \sqrt{-g}\left[ \frac{R}{16\pi G}  -\dfrac{1}{2} \sum_{i=1,2} \left(  \nabla^\mu \phi_i  \nabla_\mu \phi_i + m_i^2 \phi_i^2 \right)  \right],
}{}
together with additional higher dimension operators encoding tidal distortions,
\begin{widetext}
\eq{
\Delta S& =  \int d^4 x \sqrt{-g} \, \frac{1}{4}C_{\mu \alpha \nu \beta } C^{\rho \alpha \sigma\beta}  \sum_{i=1,2} \left( \lambda_i \phi_i^2 \delta^\mu_\rho \delta^\nu_\sigma  
+  \frac{\eta_i}{m_i^4} \nabla^\mu \nabla^\nu  \phi_i \nabla_\rho \nabla_\sigma  \phi_i  \right) .
}{eq:action}
\end{widetext}
The coefficients $\lambda_i$ and $\eta_i$ parameterize linear combinations of the mass and current quadrupoles, which we set without loss of generality to  $\lambda_1,\eta_1 \neq 0$ and $\lambda_2 = \eta_2 =0$.  
The general case is trivially obtained by symmetrizing over particle labels.  Hereafter, all variables with a $\Delta$ prefix will denote quantities linear in the tidal coefficients.

Note that higher dimension operators with more than four derivatives on the scalars require additional derivatives on the gravitational field and thus describe higher order tidal moments.  All other allowed operators can be eliminated either through field redefinitions or  Weyl tensor identities such as $C^{\mu}_{\,\, \,\alpha \beta \gamma} C^{\nu \alpha  \beta \gamma} = \frac14 g^{\mu \nu} C_{ \alpha  \beta \gamma\delta} C^{ \alpha  \beta \gamma \delta} $~\cite{Lovelock}, however they are straightforwardly included as a consistency check of the calculation.

\medskip

\begin{figure}
\begin{center}
\includegraphics[scale=.45]{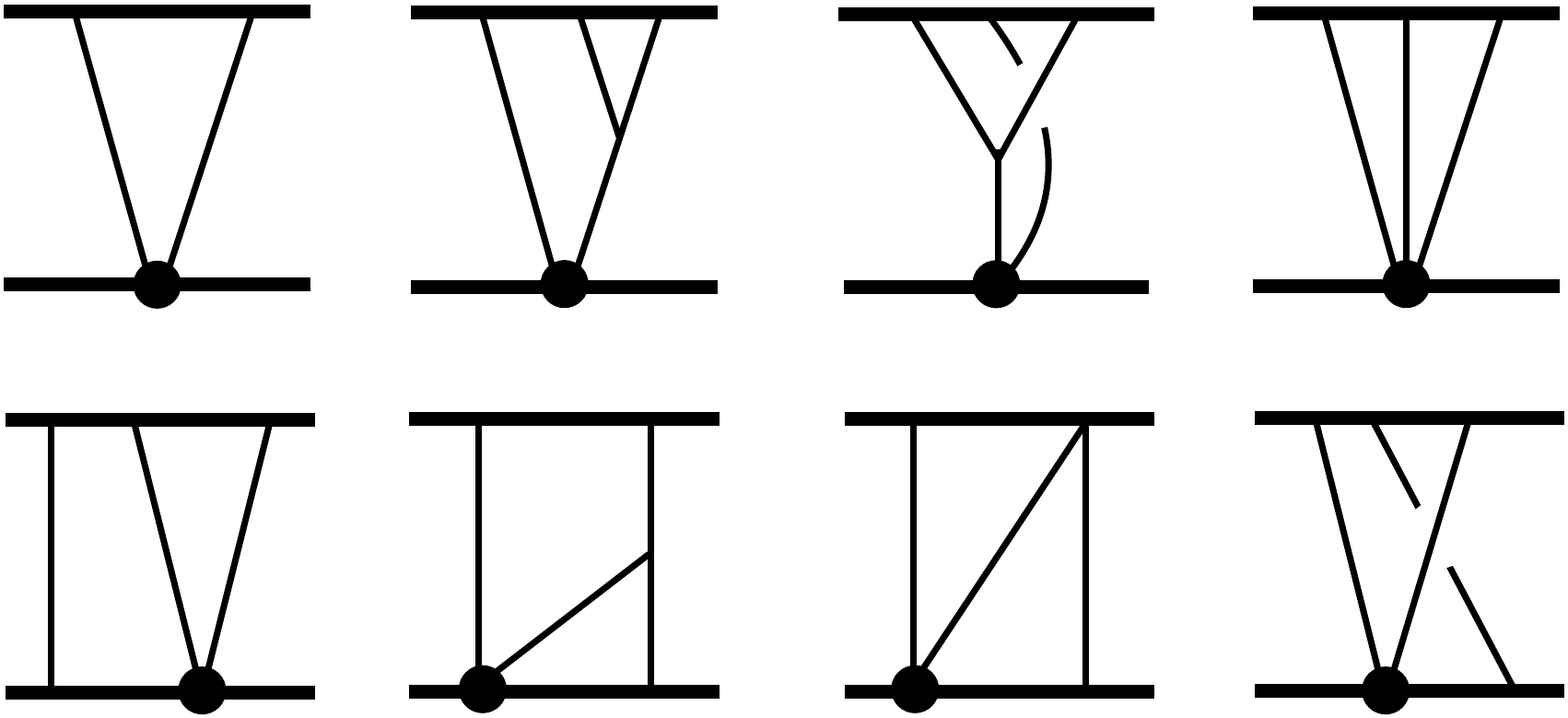}
\end{center}
\caption{Feynman diagrams for classical tidal corrections. The thick and thin lines denote massive scalars and exchanged gravitons, respectively, while black circles are tidal operator insertions.  Not shown are ``reflected'' graphs obtained by swapping the scalars, nor ``twisted'' graphs obtained by swapping the incoming and outgoing legs for one of the scalars.}
\label{fig:diagrams}
\end{figure}

\Section{Scattering Amplitudes. } 
In this section we compute the leading and next-to-leading PM tidal corrections to scattering, $\Delta M_2$ and $\Delta M_3$.    As discussed at length in \cite{3PMlong}, all diagrams with self-energy loops or contact interactions do not contribute classically.  The relevant one- and two-loop Feynman diagrams are depicted in \Fig{fig:diagrams}.  

We perform our entire calculation in the generalized graviton field basis and gauge fixing described in~\cite{3PMFeynman}, utilized previously to simplify perturbation theory~\cite{simplified} and containing deDonder gauge as a subset.  As a highly nontrivial consistency check, all gauge dependence will vanish from the physical  amplitude.  Hereafter, any gauge dependent expressions will be in deDonder gauge.
  
As described in \cite{3PMlong} the cumbersome multi-loop integrands computed using Feynman diagrams can be massively simplified by applying a procedure for classical truncation which eliminates quantum corrections at the integrand level.  Operationally, this is achieved by a series expansion in small $\epsilon$ following the replacement $ q, \ell \rightarrow \epsilon q, \epsilon \ell$ where $q$ is the four-momentum transfer and $\ell$ is any graviton loop four-momentum.   The series in $\epsilon \sim q/ p_{1,2} \sim \hbar / J$ for incoming four-momenta $p_{1,2}$ is an expansion in large angular momentum.  The classical contributions to the amplitude at ${\cal O}(G^n)$ scale as $M_n \rightarrow \epsilon^{n-3} M_n$ and $\Delta M_n \rightarrow \epsilon^{n+1} \Delta M_n$, modulo infrared divergent ``iteration'' terms which are lower order in $\epsilon$ and appear exactly in the EFT in such a way that cancels in the matching to the Hamiltonian.

At one loop there is a single ``triangle'' Feynman diagram, shown in \Fig{fig:diagrams}, that survives classical truncation.  The corresponding integrand is
\eq{
\Delta I_2 &= -  {32 G^2 \pi^2 m_2^4 \mbf{q}^4 \over \ell^2 (\ell +q)^2 (\ell^2 + 2 p_2 \cdot \ell) } \bigg[ 4\lambda_1 +{\eta_1 \over 2}  \bigg( (1-2\sigma^2)^2  \\
&\quad -   { 4(p_1 \cdot \ell)^2 \over m_1^2 \mbf{q}^2 }(1-4\sigma^2) + {8 (p_1 \cdot \ell)^4 \over m_1^4 \mbf{q}^4 }\bigg) \bigg] \,,
}{}
where $\sigma = -p_1 \cdot p_2 /m_1 m_2$. 
We then evaluate the integral $\Delta M_2 = \int \! {d^4 \ell \over (2\pi)^4} \, \Delta I_2$ via standard relativistic methods or via the non-relativistic approach in \cite{3PMlong}, obtaining
\eq{
\Delta M_2(\mbf{p}, \mbf{q})& = G^2 \pi^2  |\mbf{q}|^3 m_2^3 \left[ 4\lambda_1   + {\eta_1 \over 32} (11-30\sigma^2+35\sigma^4)\right],
}{eq:dM2}
where $\mbf{p}$ and $\mbf{q}$ are the center-of-mass three-momentum and three-momentum transfer, respectively.
 
At two loops, the calculation is substantially more complicated due to the proliferation of Feynman diagrams.  We refrain from presenting them here explicitly but include them as a supplemental attachment containing the classically truncated integrands.    We then apply the exact same integration method discussed at length in general and in examples in~\cite{3PMlong}.  In this method, the energy components of loop integrals are first localized via residues to matter poles in the potential region of the loop graviton momenta.  Afterwards, the remaining spatial integrals are expanded to very high order in velocity.  Each order then contributes a ``bubble'' integral which is evaluated via standard formulas in dimensional regularization~\cite{Smirnov}. Applying these methods to our two-loop integrands, we obtain integrated expressions up to ${\cal O}(\mbf{p}^{16})$ in the velocity expansion.  

Again following~\cite{3PMlong}, the evaluated integrals exhibit regular patterns which can be resummed to all orders in velocity into a set of simple basis kinematic functions. Resummation is possible because the velocity expansion is done only at the last step such that the integrands have vestiges of Lorentz invariance, which, together with dimensional analysis and the classical limit, imposes strong constraints on the possible momentum-dependent structures appearing. In particular, the following simple power counting argument shows that new momentum-dependent structures saturate at ${\cal O} (\mbf{p}^8)$.

Consider the scattering contribution from $\eta_i$. The associated amplitude is dimensionless and proportional to $G^3 \eta_i / m_i^4$, so the remaining kinematic dependence has mass dimension twelve. With a two-loop integral measure and at most seven propagators, the integrand numerator thus has mass dimension eighteen at most. A classical contribution requires seven powers of this mass dimension attributed to loop momenta $\ell$ with the rest attributed to external momenta $p$. This all implies a numerator with schematic structure $p^4 (p \cdot \ell)^7$, where $p^4$ is effectively a constant that can be factored out of the integral. Since the resulting integral produces at most a rank seven tensor, no new momentum-dependent structures can arise beyond ${\cal O} (\mbf{p}^8)$.

In the end, this procedure produces the following two-loop tidal correction to the scattering amplitude,
\begin{widetext}
\eq{
\Delta  M_3(\mbf{p}, \mbf{q})& =   G^3 \pi \,   \mbf{q}^4 \ln |\mbf{q}| \, m_2^3   \Bigg[   
  4 \lambda_1 \Bigg( {8  m_2  \over 5} 
   -  { m_1 \sigma (5-2\sigma^2) \over (\sigma^2 - 1)^2} + { 6 m_1  \, {\sinh}^{-1} \sqrt{\sigma - 1 \over 2} \over (\sigma^2- 1)^{5/2}}  \Bigg) + \eta_1 \Bigg( {\, m_2 (305 - 363 \sigma^2 - 110 \sigma^4) \over 560}   \\
   & \quad -{ m_1 \sigma (5401 - 195 \sigma^2 -94 \sigma^4 ) \over 80 } - {  m_1 \sigma (673 + 2168 \sigma^2 ) \over 2(\sigma^2- 1)^{2} }  + {3  m_1 \left(33+ 474 \sigma^2 + 440 \sigma^4  \right) {\sinh}^{-1} \sqrt{\sigma - 1 \over 2}  \over (\sigma^2 - 1)^{5/2} }  \Bigg) \\
& \quad  + 2(1-2\sigma^2) \left[ 4 \lambda_1 + {\eta_1 \over 32} (11-30 \sigma^2 + 35 \sigma^4) \right]  {E (E_2 -m_2) \over m_2 (\sigma^2 - 1)}    \Bigg] + \frac{1}{E} 
\int \frac{d^3{\boldsymbol{\ell}}}{(2\pi)^3} 
\, \frac{M_1(\mbf{p}, \mbf{\ell}) \Delta M_2(\mbf{p}, \boldsymbol{\ell} + \boldsymbol{q})}{ (\mbf{\ell}^2 + 2 \mbf{p} \mbf{\ell})} \,,
}{eq:M_full}
\end{widetext}
where $E = E_1 +E_2$ is the total energy and the nonrelativistic normalization $1/(4E_1E_2)$ has {\it not} been included. 

\Eq{eq:M_full} is reminiscent of the ${\cal O}(G^3)$ scattering amplitude for point-like  objects~\cite{3PM,3PMlong}. 
In the $m_1 \ll m_2$ expansion, the first term within each round bracket is dominant and all remaining terms are captured at next-to-leading order.  This accords with the expected mass dependence of the classical scattering angle~\cite{Vines:2018gqi,Damour:2019lcq}, implying that the full ${\cal O}(G^3)$ dynamics are accessible from a first order self-force calculation.  By the same logic, the ${\cal O}(G^2)$ amplitude is completely fixed by the test-particle limit.

As similarly observed for point-like compact objects \cite{3PM,3PMlong}, the final line in \Eq{eq:M_full} contains both finite and infrared divergent contributions from the iteration of the one-loop tidal contribution $\Delta M_2$ with the tree-level amplitude ${M}_1(\mbf{p},\mbf{q}) = -{16\pi G m_1^2 m_2^2 (1-2\sigma^2)/}{\mbf{q}^2}$.  

As another check of our resummation, we take $\Delta M_3$ in \Eq{eq:M_full} and reweight each kinematically-independent term by a free coefficient.  We find that this ansatz is uniquely fixed to \Eq{eq:M_full} after taking as input our explicitly integrated amplitude at ${\cal O}(\mbf{p}^{10})$.  Consequently, the match of our integrated results to \Eq{eq:M_full} at ${\cal O}(\mbf{p}^{16})$ is a highly nontrivial consistency check. 

\medskip

\Section{Matching.} Following the approach of \cite{2PM}, we construct an isotropic gauge EFT Hamiltonian, including point-particle contributions
\eq{
H^{\rm EFT}(\mbf{p},\mbf{r})&= \sum_{i=1,2} \sqrt{\mbf{p}^2 + m_i^2 } + \sum_{n=1}^\infty \frac{G^n c_n(\mbf{p}^2)}{|\mbf{r}|^n}\,,
}{}
as well as tidal corrections
\eq{
\Delta H^{\rm EFT}(\mbf{p},\mbf{r}) &=  \sum_{n=2}^\infty \frac{G^n \Delta c_n(\mbf{p}^2)}{|\mbf{r}|^{n+4}}\,,
}{}
where $\mbf{p}$ and $\mbf{r}$ are the center of mass momentum and distance between bodies.  Here $c_n(\mbf{p}^2)$ and $\Delta c_n(\mbf{p}^2)$ dictate the momentum-dependent interactions at zeroth and first order in the tidal coefficients. Explicit expressions for $c_n(\mbf{p}^2)$ can be found in Eq.~(10.10) of Ref.~\cite{3PMlong}. 

The EFT amplitudes, $M^{\rm EFT}$  and $\Delta M^{\rm EFT}$, can be trivially computed via Feynman diagrams within the framework of \cite{2PM} that was used to obtain all of the results in \cite{3PMlong}. There it was also observed by explicit calculation that the EFT amplitude in position space is exactly proportional to the local center of mass momentum squared $\mbf{p}^2_{\rm loc} (\mbf{p},\mbf{r})$ as a function of the incoming asymptotic momentum $\mbf{p}$ and the separation $\mbf{r}$ between bodies.  Consequently, the EFT amplitude can be extracted {\it algebraically} from the classical equations of motion, as was later proven in \cite{B2B,LSE}.  
Applying this simpler procedure, we obtain
\eq{
\Delta M_2^{\rm EFT} (\mbf{p},\mbf{q})&=  - \frac{G^2 \pi^2 | \mbf{q}|^3 \Delta c_2}{12} 
}{}
for the ${\cal O}(G^2)$ tidal correction to the amplitude and
\begin{widetext}
\eq{
\Delta M_3^{\rm EFT} (\mbf{p},\mbf{q})&=  \frac{G^3 \pi \mbf{q}^4 \ln |\mbf q|  }{30}  \left[ \Delta c_3 - {(1-3 \xi) c_1 \Delta c_2 \over E \xi} - 2 E \xi (c_1\Delta c_2)'  \right] + 4 E \xi  \int \frac{d^3{\boldsymbol{\ell}}}{(2\pi)^3}  \, \frac{M_1^{\rm EFT}(\mbf{p},\mbf{\ell}) \Delta M_2^{\rm EFT}(\mbf{p}, \boldsymbol{\ell} + \boldsymbol{q}) }{ (\mbf{\ell}^2 + 2 \mbf{p} \mbf{\ell})} 
}{}
\end{widetext}
at ${\cal O}(G^3)$, where $\xi = E_1 E_2/(E_1+E_2)^2$ and primed differentiation is performed with respect to $\mbf p^2$. Here we have written the infrared divergent contribution in terms of the iteration of the one-loop ${\cal O}(G^2)$ EFT amplitude $\Delta M_2^{\rm EFT}$ together with the ${\cal O}(G)$ point-particle EFT amplitude $M_1^{\rm EFT}(\mbf{p},\mbf{q}) = -4 \pi G c_1(\mbf{p}^2)/\mbf{q}^2$. Equating these EFT amplitudes to those in the full theory with nonrelativistic normalization factor $1/4E_1E_2$, we obtain the tidal corrections to the conservative Hamiltonian,
\begin{widetext}
\eq{
\Delta c_2 &=-{3m_2^3 \over E^2 \xi} \left[ 4 \lambda_1 + {\eta_1 \over 32} (11-30 \sigma^2 + 35 \sigma^4) \right] \,, \\
\Delta c_3 &= {15 m_2^3 \over 2 E^2 \xi}     \Bigg[   
  4 \lambda_1 \Bigg( {8  m_2  \over 5} 
   -  { m_1 \sigma (5-2\sigma^2) \over (\sigma^2 - 1)^2} + { 6 m_1  \, {\sinh}^{-1} \sqrt{\sigma - 1 \over 2} \over (\sigma^2- 1)^{5/2}}  \Bigg) + \eta_1 \Bigg( {\, m_2 (305 - 363 \sigma^2 - 110 \sigma^4) \over 560}   \\
   & \quad -{ m_1 \sigma (5401 - 195 \sigma^2 -94 \sigma^4 ) \over 80 } - {  m_1 \sigma (673 + 2168 \sigma^2 ) \over 2(\sigma^2- 1)^{2} }  + {3  m_1 \left(33+ 474 \sigma^2 + 440 \sigma^4  \right) {\sinh}^{-1} \sqrt{\sigma - 1 \over 2}  \over (\sigma^2 - 1)^{5/2} }  \Bigg) \\
& \quad  + 2(1-2\sigma^2) \left[ 4 \lambda_1 + {\eta_1 \over 32} (11-30 \sigma^2 + 35 \sigma^4) \right]  {E (E_2 -m_2) \over m_2 (\sigma^2 - 1)}    \Bigg] \\
&\quad + {3 \nu m_2^3 \over m \gamma^5 \xi^3} \bigg[ \nu(1-\xi)(1-2\sigma^2)\left[ 4 \lambda_1 + {\eta_1 \over 32} (11-30 \sigma^2 + 35 \sigma^4) \right] + 4\gamma^2 \xi \sigma \left[ 4 \lambda_1 + {\eta_1 \over 32} (26- 95 \sigma^2 + 105 \sigma^4) \right] \bigg] \,,
}{eq:ds}
\end{widetext}
where $\gamma = E/m$ and $m=(m_1+m_2)$.
As anticipated, the infrared divergent contributions to $\Delta M$ and $\Delta M^{\rm EFT}$ cancel exactly, which is itself a consistency check.

\medskip

\Section{Worldline Action.}
It will be useful to recast our expressions in terms of the standard notation for the tidal moments in the worldline formalism.  The action is given by the point-particle contribution $S^{\rm WL} = -\sum_{i=1,2} m_i\int  d\tau_i  $ together with the leading tidal corrections,
\eq{
\Delta S^{\rm WL} &=  \sum_{i=1,2} \int d\tau_i \left[ \frac{\mu^{(2)}_i }{4} \left({E}^{i}_{\alpha  \beta}\right)^2+ \frac{2\sigma^{(2)}_i}{3} \left({ B}^{i}_{\alpha  \beta} \right)^2 \right]\, .
}{eq:WL_coeffs}
Here the mass and current quadrupole moments are parameterized in the conventions of Refs.~\cite{EOBTidal,Henry:2019xhg}.  The gravito-electric and gravito-magnetic tensors, ${ E}^{i}_{\alpha  \beta}$ and ${ B}^{i}_{\alpha  \beta}$, are related to the Weyl curvature invariants evaluated on each worldline by 
$(C_{\alpha\beta\gamma\delta})^2= 8 ({ E}^{i}_{\alpha  \beta})^2 -8( { B}^i_{\alpha  \beta})^2 $ and $(u_i^\mu u_i^\nu C_{\mu \alpha \nu \beta})^2  = ({ E}^{i}_{\alpha  \beta})^2$, where  the four-velocity $u_i$ of each particle satisfies $u_i^2 = -1$.  We can then relate the tidal coefficients in \Eq{eq:action} to those in \Eq{eq:WL_coeffs},
\eq{
\frac{ \lambda_i}{ m_i} = - \frac{\sigma^{(2)}_i}{3}  \qquad \textrm{and} \qquad \frac{\eta_i}{m_i} =\mu^{(2)}_i + \frac{8\sigma^{(2)}_i}{3} \,,
}{eq:WL_relations}
allowing us to make contact with results derived in the worldline formalism.  For example, our result for the leading order tidal correction described by $\Delta c_2$ exactly matches the result in Eq.~(5.13) of Ref.~\cite{PMGR} which was derived using a PM worldline effective theory.

\medskip

\Section{Hamiltonian and Binding Energy.}  Our Hamiltonian is consistent with all existing results at the relevant overlapping 2PN accuracy.  To show this we transform the tidal Lagrangian in Eq.~(5.4) of~\cite{Henry:2019xhg} to a Hamiltonian, as usual taking special care to eliminate acceleration terms and account for the induced shift in coordinates.   Again using the EFT methods of~\cite{2PM}, we then compute the tidal corrections to scattering from this Hamiltonian and find exact agreement with the 2PN terms in \Eq{eq:dM2} and \Eq{eq:M_full}, i.e.~the terms at ${\cal O}(G^2 v^4)$ and ${\cal O}(G^3 v^2)$ and lower. We thus conclude that the 2PN truncation of our tidal Hamiltonian in \Eq{eq:ds} is gauge equivalent to existing results.

As an additional albeit redundant check, we use \Eq{eq:ds} to compute the tidal corrections to the 1PN circular binding energy and find exact agreement with Eq.~(6.5) of ~\cite{Henry:2019xhg}.

\medskip

\Section{Scattering Angle.} Another gauge invariant physical quantity we can compute is the conservative contribution to the classical scattering angle. In Ref.~\cite{3PM}, it was observed that this is directly related to the finite parts of the scattering amplitude, and this structure has now been understood to all orders~\cite{B2B,LSE}. The tidal correction to the scattering angle through next-to-leading order is
\eq{
2 \pi \Delta \chi &= {45 \mbf{p}^4 \Delta{\widetilde M}_2  \over 4 E  J^6 }  - {96 \mbf{p}^5  \Delta{\widetilde M}_3  \over E  J^7 } + {12 \mbf{p}^3  {\widetilde { M}}_1 \Delta{\widetilde M}_2  \over E^2 J^7 \pi^2  }\,, 
}{}
where tilded quantities denote finite parts of the corresponding relativistically normalized amplitude contributions with the $\mbf{q}$ dependence stripped off, i.e.~$\Delta {\widetilde M}_2= \Delta {M}_2 /|\mbf{q}|^3$, $\Delta {\widetilde M}_3= \Delta {M}_3 /( \mbf{q}^4 \ln |\mbf{q}|)$ and ${\widetilde M}_1=  {M}_1  \mbf{q}^2$, where $M_1$ is the point particle amplitude from Born exchange defined earlier. We have checked that the ${\cal O}(J^{-6})$ contribution agrees with Eq.~(6.2) of~\cite{Bini:2020flp}. 

\medskip

\Section{Test-Particle Limit}. Our expressions are valid in the test-particle limit.  Consider the case of a neutron star orbiting a supermassive black hole.   In the strict $m_1 \ll m_2$ limit, particle 1 is effectively a non-minimally coupled test mass residing on a  background Schwarzschild spacetime sourced by particle 2.  Following the approach of \cite{Steinhoff:2016rfi}, the geodesic trajectory for particle 1 is dictated by the mass shell condition on its  four-momentum $p$,
\eq{
0= p^\mu p_{\mu} + m_1^2-\frac{\lambda_1}{2} C_{\alpha\beta\gamma\delta}^2- \frac{\eta_1}{2m_1^4} (p^\mu p^\nu C_{\mu \alpha \nu \beta})^2,
}{}
where all metric contractions are performed with the Schwarzschild metric $g_{\mu\nu}$, taken here to be in isotropic coordinates.
We then identify the energy component $p_0 = H^{\rm Sch}+ \Delta H^{\rm Sch}$ with the test-particle Hamiltonian including point-particle contributions \cite{WS93}
\eq{
H^{\rm Sch} (\mbf{p},\mbf{r}) =  f_- (\mbf{r}) f_+ (\mbf{r})^{-3} (\mbf{p}^2+ f_+(\mbf{r})^{4} m_1^2 )^{1/2} \,,
}{eq:H_test}
as well as corrections linear in the tidal coefficients,
\begin{widetext}
\eq{
\Delta H^{\rm Sch} (\mbf{p},\mbf{r}) &=  3\lambda_1 \left[ -\frac{R^2}{\mbf{r}^6 E_1} + \frac{R^3 ( 6 \mbf p^2 + 7 m_1^2)}{2 |\mbf{r}|^7 E_1^3} \right]   + \frac{3\eta_1}{8} \left[  -  \frac{R^2 \left( 1 + 3 \mbf{v}_\perp^2 + 3 \mbf{v}_\perp^4 \right)}{\mbf{r}^6  E_1}  \right.  \\
& \quad \left. +    \frac{R^3 \left(  \mbf{p}^2 \left( 6 + 24 \mbf{v}_\perp^2 + 30 \mbf{v}_\perp^4 \right)
    +m_1^2 \left( 7 + 27 \mbf{v}_\perp^2 +33 \mbf{v}_\perp^4 \right) \right)}{2|\mbf{r}|^7 E_1^3} \right] + {\cal O}( \mbf{r}^{-8}) \,,
}{eq:dH_test}
\end{widetext}
where $f_{\pm}(\mbf{r}) = 1\pm \frac{R}{4 |\mbf{r}|}$, $R = 2Gm_2$ is the Schwarzschild radius, and $\mbf{v}_\perp^2 = \bigr( \mbf{p}^2 - \tfrac{(\mbf p  \cdot \mbf r)^2 }{ \mbf r^2} \bigr)/m_1^2$.
Note that for $m_1 \ll m_2$, \Eq{eq:H_test} is valid to all orders in $\bm p$ and $\bm r$ while \Eq{eq:dH_test} is truncated at ${\cal O}( |\mbf{r}|^{-7})$ for the sake of brevity. 

It is easy to see that the terms in $\Delta H^{\rm Sch}$ proportional to $\lambda_1$ at ${\cal O}(\mbf{r}^{-6})$ and ${\cal O}(|\mbf{r}|^{-7})$ coincide exactly with the $m_1\ll m_2$ limit of $\Delta c_2$ and $\Delta c_3$ in \Eq{eq:ds}.   On the other hand, a comparison of the $\eta_1$ corrections is not so straightforward since the relevant terms in $\Delta H^{\rm Sch}$ depend on $\mbf{p} \cdot \mbf{r}$ and thus depart from the isotropic gauge of \Eq{eq:ds}. Hence, a proper comparison requires constructing a canonical transformation between gauges or, alternatively, computing a physical, gauge invariant quantity such as the scattering amplitude. Using the EFT approach of \cite{2PM}, we compute the $\eta_1$ tidal corrections to the scattering amplitude and find an exact match to the $m_1 \ll m_2$ limit of \Eq{eq:dM2} and \Eq{eq:M_full} after including non-relativistic normalization. This match between scattering amplitudes implies that the test-particle limit of our Hamiltonian is gauge equivalent to \Eq{eq:dH_test}.

\medskip

\Section{Discussion.} We have presented the first ever calculation of tidal corrections to the conservative Hamiltonian for spinless compact objects at next-to-leading order in the PM expansion. These dynamics are extracted from the two-loop ${\cal O}(G^3)$ scattering amplitude at linear order in the mass and current quadrupole moments.  

Our expressions pass many checks.  Still, it would be interesting to verify them with traditional methods, e.g.~as was done for the point particle 3PM Hamiltonian \cite{3PM, 3PMlong}  
 at 5PN and 6PN via self-force theory~\cite{BDG} and PN perturbation theory~\cite{Blumlein:2020znm}.  Also valuable would be a comparison of our results against other approaches like numerical relativity and effective one-body formalism, as was done in \cite{Antonelli:2019ytb} for the case of the binary black hole inspiral.

\medskip

\Section{Acknowledgements.}
 We are grateful to Zvi Bern, Luc Blanchet, Thibault Damour, Walter Goldberger, Ira Rothstein, Jan Steinhoff, and Justin Vines for comments on this manuscript.
C.C. and M.P.S. are supported by the DOE under grant no.~DE-SC0011632 and by the Walter Burke Institute for Theoretical Physics. The calculations here
used the computer algebra system \texttt{Mathematica}~\cite{Mathematica} in combination with \texttt{FeynCalc}~\cite{FeynCalc} and \texttt{xAct}~\cite{xAct}, as well as the On-Line Encyclopedia of Integer Sequences~\cite{OEIS} and the Hoffman2 Cluster at the Institute for Digital Research and Education at UCLA.

%%%%%%%%%%%%%%%%%%%%

\end{document}